\documentclass[12pt]{article}

\usepackage{cite}
\usepackage{epsfig}
\usepackage{amsmath}
\usepackage{array}
\usepackage{pstricks}
\usepackage{bbold}

\def\mathswitch#1{\relax\ifmmode#1\else$#1$\fi}
\def\mathswitchr#1{\relax\ifmmode{\mathrm{#1}}\else$\mathrm{#1}$\fi}
\newcommand{\PW}{\mathswitchr W}
\newcommand{\PZ}{\mathswitchr Z}
\newcommand{\PH}{\mathswitchr H}

\newcommand{\Pb}{\mathswitchr b}
\newcommand{\Pt}{\mathswitchr t}

\newcommand{\MW}{\mathswitch {M_\PW}}
\newcommand{\MZ}{\mathswitch {M_\PZ}}
\newcommand{\MH}{\mathswitch {M_\PH}}

\newcommand{\mb}{\mathswitch {m_\Pb}}
\newcommand{\mt}{\mathswitch {m_\Pt}}

\newcommand{\mz}{\mathswitch {\overline{M}_\PZ}}

\newcommand{\gz}{\mathswitch {\overline{\Gamma}_\PZ}}
\newcommand{\as}{\alpha_{\mathrm s}}

\newcommand{\gev}{\,\, \mathrm{GeV}}
\newcommand{\mev}{\,\, \mathrm{MeV}}

\newcommand{\OO}{{\mathcal O}}

\newcommand{\mycaption}[1]{\caption{\sl #1}}

\begin{document}
\thispagestyle{empty}

\def\thefootnote{\fnsymbol{footnote}}

\begin{flushright}
\end{flushright}

\vspace{1cm}

\begin{center}

{\Large\sc {\bf Two-loop fermionic electroweak corrections\\[1ex] to 
the $Z$-boson width and production rate}}
\\[3.5em]
{\large\sc
Ayres~Freitas
}

\vspace*{1cm}

{\sl
Pittsburgh Particle-physics Astro-physics \& Cosmology Center
(PITT-PACC),\\ Department of Physics \& Astronomy, University of Pittsburgh,
Pittsburgh, PA 15260, USA
}

\end{center}

\vspace*{2.5cm}

\begin{abstract}
Improved predictions for the $Z$-boson decay width and the hadronic $Z$-peak
cross-section within the Standard Model are presented, based on a complete 
calculation of electroweak two-loop corrections with
closed fermion loops. Compared to previous partial results, the
predictions for the $Z$ width and hadronic cross-section shift by about 0.6~MeV
and 0.004~nb, respectively. Compact parametrization formulae are provided, which
approximate the full results to better than 4 ppm.
\end{abstract}

\setcounter{page}{0}
\setcounter{footnote}{0}

\newpage


The recent discovery of the Higgs boson \cite{higgs} has been a tremendous
success of the Standard Model (SM). Remarkably, the observed mass of the Higgs
boson, $\MH$, agrees very well with the value that has been predicted many years
earlier from electroweak (EW) precision observables, which are quantities that
have been measured with very high accuracy at lower energies, and that can be
calculated comparably precisely within the SM. Besides predicting the Higgs
mass, global fits to all available electroweak precision observables 
are crucial for testing the SM at the quantum
level and constraining new physics (see Ref.~\cite{lepz,gfit,xfit} for recent
examples).

On the theory side, these precision tests rely on calculations of radiative
corrections for the relevant observables. At the current level of precision,
two-loop EW and higher-order QCD corrections are numerically important
\cite{mwlong,swlept}.
Complete two-loop contributions are known for the prediction of
the mass of the $W$ boson, $\MW$ \cite{qcd2,mw,mwlong,mwtot}, and the leptonic
effective weak mixing angle, $\sin^2 \theta_{\rm eff}^\ell$
\cite{qcd2,swlept,swlept1,swlept2}, which is related to the ratio of vector and
axial-vector couplings of the $Z$ boson to leptons. Furthermore, the leading
three- and four-loop corrections to these observables in the limit of large values of the top-quark
mass, $\mt$, have been obtained. These are EW contributions of order
$\OO(\alpha^3\mt^6)$ \cite{mt6} and mixed EW/QCD terms of $\OO(\alpha\as^2)$
\cite{qcd3}, $\OO(\alpha^2\as\mt^4)$ \cite{mt6}, and $\OO(\alpha\as^3\mt^2)$ 
\cite{qcd4}. Similarly precise results are available for the effective
weak mixing angle of quarks \cite{swlept2,swbb} and the ratio of the $Z$-boson
partial widths into $b\bar{b}$ and all hadronic final states, $R_b
\equiv \Gamma_{Z \to b\bar{b}}/\Gamma_{Z \to \text{had.}}$ \cite{rb}, except that
the so-called \emph{bosonic} EW two-loop corrections stemming from
diagrams without closed fermion loops are not known for these quantities.
However, detailed analyses and experience from the calculation of $\MW$ and
$\sin^2 \theta_{\rm eff}^\ell$ indicates that these are small.
Most of the published results have been implemented into the public code
\textsc{ZFitter} \cite{zfitter} and several private packages \cite{xfit,gfitter}.

However, for the decay width and production cross-section of the $Z$ boson, so
far only approximate results for the EW two-loop corrections have been
calculated in a large-$\mt$ expansion, including the next-to-leading order
$\OO(\alpha^2\mt^2)$ \cite{ewmt2} for leptonic final states and quarks of the
first two generations, while for the $Z\to b\bar{b}$ partial width merely the
leading $\OO(\alpha^2\mt^4)$ coefficient is known \cite{ewmt4}. It was estimated
that the missing EW two-loop corrections lead to an uncertainty of several
MeV for the prediction of the $Z$ width \cite{snowmass}, 
which is comparable to the experimental error, but has not been properly
accounted for in the global SM fits.
In this letter, this gap is filled by presenting the complete \emph{fermionic} EW two-loop corrections
($i.\,e.$ from diagrams with one or two closed fermion loops) for the $Z$-boson
width and production rate. With these new results, the theoretical
uncertainty from missing higher-order contributions in electroweak fits will be
significantly reduced.


\vspace{\bigskipamount}
The width of the $Z$ boson is defined through the imaginary part of the complex
pole $s_0 \equiv \mz^2 - i\mz \gz$ 
of the propagator
\begin{equation} 
[s - \mz^2 + \Sigma_\PZ(s)]^{-1}, \label{prop}
\end{equation}
where $\Sigma_\PZ(s)$ is the transverse part of the $Z$ self-energy, resulting in
\begin{equation}
\gz = \frac{1}{\mz} \text{Im}\,\Sigma_\PZ(s_0).
\label{gz}
\end{equation}
This definition is consistent and gauge-invariant too all orders \cite{prop}.
Expanding eq.~\eqref{gz} up to next-to-next-to-leading
order in the electroweak coupling and using the optical theorem leads to (see Ref.~\cite{long} for details)
\begin{align}
\gz &= \sum_f \overline{\Gamma}_f\,, \qquad
\overline{\Gamma}_f = \frac{N_c\mz}{12\pi} \Bigl [
 {\cal R}_{\rm V}^f F_{\rm V}^f + {\cal R}_{\rm A}^f F_{\rm A}^f \Bigr ]_{s=\mz^2} 
 \;, \label{Gz} \displaybreak[0] \\
F_{\rm V}^f &= v_{f(0)}^2
 \bigl [1-\text{Re}\,\Sigma'_{\PZ(1)}-\text{Re}\,\Sigma'_{\PZ(2)}
  + (\text{Re}\,\Sigma'_{\PZ(1)})^2 \bigr ] 
 + 2 \,\text{Re}\, (v_{f(0)}v_{f(1)})\bigl [1-\text{Re}\,\Sigma'_{\PZ(1)} \bigr ] 
\nonumber \\[.5ex] 
 &\quad + 2 \,\text{Re}\, (v_{f(0)}v_{f(2)}) + |v_{f(1)}|^2
 - \tfrac{1}{2}\mz \gz v_{f(0)}^2
 \;\text{Im}\,\Sigma''_{\PZ(1)}\;, \label{Fv}
\end{align}
where $v_f$ is the effective vector $Zf\bar{f}$
coupling, which includes EW vertex corrections and 
$Z$--$\gamma$ mixing contributions, and $F_A^f$ is defined similarly in terms of the
axial-vector coupling $a_f$. The subscripts in brackets indicate the loop order and
$\Sigma'_\PZ$ is the derivative of $\Sigma_\PZ$. The radiator functions ${\cal
R}_{\rm V,A}$ take into account final-state QED and QCD radiation and are known
up to ${\cal O}(\as^4)$ in the limit of massless quarks and ${\cal O}(\as^3)$
for the kinematic mass corrections \cite{rad}.

Note that the mass and width defined through the complex pole of \eqref{prop}
differ from the experimentally reported values, $\MZ$ and $\Gamma_\PZ$, since the
latter have been obtained using a Breit-Wigner formula with a running width. The
two are related via
\begin{equation}
\textstyle
\mz = \MZ\big/\sqrt{1+\Gamma_\PZ^2/\MZ^2}\,, \qquad
\gz = \Gamma_\PZ\big/\sqrt{1+\Gamma_\PZ^2/\MZ^2}\,,
\end{equation}
amounting to $\mz \approx \MZ - 34\mev$ and $\gz \approx \Gamma_\PZ - 0.9\mev$.

Now let us turn to the $Z$-boson cross-section.
After subtracting contributions from photon exchange and box diagrams, the
amplitude for $e^+e^- \to f\bar{f}$ can be written as an expansion about the
complex pole,
\begin{equation}
{\cal A}_\PZ(s) = \frac{R}{s-s_0} + S + (s-s_0)S' + \dots
\label{amp}
\end{equation}
Instead of the total cross-section, it is customary to use the hadronic peak
cross-section defined as
\begin{equation}
\sigma^0_{\rm had} = \frac{1}{64\pi^2\MZ^2}
 \sum_{f=u,d,c,s,b}\int d\Omega\;\bigl|{\cal A}_\PZ(\MZ^2) 
 \bigr|^2 .
\end{equation}
Starting from \eqref{amp}, an explicit calculation yields
\begin{equation}
\sigma^0_{\rm had} = \sum_{f=u,d,c,s,b}
 \frac{12\pi}{\mz^2}\,\frac{\overline{\Gamma}_e\overline{\Gamma}_f}{\gz^2}
 (1+\delta X)\,,
\end{equation} 
where $\delta X$ occurs first at two-loop level \cite{Grassi:2000dz} and is given
by\footnote{Note that eq.~\eqref{dx} differs from the expression shown in
Ref.~\cite{Grassi:2000dz} since the non-resonant terms in eq.~\eqref{amp} were not included
there. See Ref.~\cite{long} for details.}
\begin{equation}
\delta X_{(2)} = -(\text{Im}\,\Sigma'_{\PZ(1)})^2 - 2\gz\mz \;\text{Im}\,\Sigma''_{\PZ(1)}
\label{dx}\,.
\end{equation} 
The calculation of the ${\cal O}(\alpha^2)$ corrections to $\gz$ and
$\sigma^0_{\rm had}$ was carried out as follows: Diagrams for the form factors
$v_{f(n)}$ and $a_{f(n)}$ were generated with {\sc FeynArts 3.3} \cite{feynarts}. 
For the renormalization the
on-shell scheme has been used, as described in Ref.~\cite{mwlong}.
Two-loop self-energy integrals and vertex integrals with sub-loop self-energy
bubbles have been evaluated with the method illustrated in section~3.2 of
Ref.~\cite{swlept2}, while for vertex diagrams with sub-loop triangles the
technique of Ref.~\cite{sub} has been employed.
As a cross-check, the results for $\sin^2 \theta_{\rm eff}^\ell$ \cite{swlept},
$\sin^2 \theta_{\rm eff}^b$ \cite{swbb}, and $R_b$ \cite{rb} have been
reproduced using the effective couplings $v_{f(n)}$ and $a_{f(n)}$, and very
good agreement within theory uncertainties has been obtained.


\vspace{\bigskipamount}
For the presentation of numerical results, the one- and two-loop EW
corrections have been combined with virtual loop corrections of order
$\OO(\alpha\as)$ \cite{qcd2}, which have been re-computed for this work, and 
partial higher-order corrections proportional to $\alpha_\Pt \equiv \alpha
\mt^2$, of order $\OO(\alpha_\Pt\as^2)$ \cite{qcd3}, 
$\OO(\alpha_\Pt^2\as)$, $\OO(\alpha_\Pt^3)$ \cite{mt6}, and $\OO(\alpha_\Pt\as^3)$ 
\cite{qcd4}. Final-state QED and QCD radiation has been included via the
radiator functions ${\cal R}_{\rm V,A}$ as described after eq.~\eqref{Fv}. 
However, the factorization between 
EW corrections ($F_{\rm V,A}$) and final-state radiation (${\cal
R}_{\rm V,A}$) in eq.~\eqref{Gz} is not exact, but additional non-factorizable
contributions appear first at $\OO(\alpha\as)$ \cite{nfact,nfactb}. All fermion masses
except for $\mt$ have been neglected in the EW two-loop corrections,
but a finite $b$ quark mass has been retained in the $\OO(\alpha)$ and
$\OO(\alpha\as)$ contributions, and non-zero bottom, charm and tau masses are
included in the radiators ${\cal R}_{\rm V,A}$.

The final combined result is evaluated as a perturbative expansion in $\alpha$
and $\as$, rather than the Fermi constant $G_\mu$. Using the input parameters
in Tab.~\ref{tab:input}, except $G_\mu$, the size of various loop contributions is
shown in Tab.~\ref{tab:res1}.

\begin{table}[tb]
\renewcommand{\arraystretch}{1.2}
\begin{center}
\begin{tabular}{|ll|ll|}
\hline
Parameter & Value & Parameter & Value \\
\hline \hline
$\MZ$ & 91.1876 GeV & $\mb^{\overline{\rm MS}}$ & 4.20 GeV \\
$\Gamma_\PZ$ & 2.4952 GeV & $m_{\rm c}^{\overline{\rm MS}}$ & 1.275 GeV \\
$\MW$ & 80.385 GeV & $m_\tau$ & 1.777 GeV \\
$\Gamma_\PW$ & 2.085 GeV & $\Delta\alpha$ & 0.05900 \\
$\MH$ & 125.7 GeV & $\as(\MZ)$ & 0.1184 \\
$\mt$ & 173.2 GeV & $G_\mu$ & $1.16638 \times 10^{-5}$~GeV$^{-2}$ \\
\hline
\end{tabular}
\end{center}
\vspace{-2ex}
\mycaption{Input parameters used for Tabs.~\ref{tab:res1} and \ref{tab:res2}, 
from Ref.~\cite{pdg,gfit}. Here $\Delta\alpha$ is the shift in the
electromagnetic coupling due to loop corrections from leptons
\cite{dalept} and the five light quark flavors, $\Delta\alpha=\Delta\alpha_{\rm
lept}(\MZ)+\Delta\alpha^{(5)}_{\rm had}(\MZ)$.
\label{tab:input}}
\end{table}
\begin{table}[tb]
\renewcommand{\arraystretch}{1.2}
\begin{center}
\begin{tabular}{|l|r|r|}
\hline
 & $\Gamma_\PZ$ [MeV] & $\sigma^0_{\rm had}$ [pb] \\
\hline \hline
$\OO(\alpha)$ & 60.26 & $-48.85$ \\
$\OO(\alpha\as)$ & 9.11 & 3.14 \\
$\OO(\alpha_\Pt\as^2,\,\alpha_\Pt\as^3,\,\alpha^2_\Pt\as,\,\alpha_\Pt^3)$ &
 1.20 & 0.48 \\
$\OO(N_f^2\alpha^2)$ & 5.13 & $-1.03$ \\
$\OO(N_f\alpha^2)$ & 3.04 & 9.07 \\
\hline
\end{tabular}
\end{center}
\vspace{-2ex}
\mycaption{Loop contributions to $\Gamma_\PZ$ and $\sigma^0_{\rm had}$ with
fixed $\MW$ as input parameter. Here $N_f$ and $N_f^2$ refer to corrections with 
one and two closed fermion loops, respectively, and $\alpha_\Pt = \alpha\mt^2$.
In all rows the radiator functions ${\cal R}_{\rm V,A}$ with known contributions
through $\OO(\as^4)$, $\OO(\alpha^2)$ and $\OO(\alpha\as)$ are included.
\label{tab:res1}}
\end{table}

However, it is common practice not to use the $W$ mass, $\MW$, as an input
parameter, but instead to determine $\MW$ from $G_\mu$. Using the results from
Ref.~\cite{mw,mwlong,mwtot} for the computation of $\MW$, the numbers in
Tab.~\ref{tab:res2} are obtained. For comparison, also shown are the
corresponding results based on the approximation of the EW two-loop corrections
for large values of $\mt$ \cite{ewmt4,ewmt2}\footnote{These numbers have been
kindly supplied by S.~Mishima based on the work in Ref.~\cite{xfit}.}. The
difference illustrates the impact of the full fermionic two-loop corrections
beyond the large-$\mt$ approximation, which can be seen to be of moderate size,
about 0.6~MeV for $\Gamma_\PZ$ and about 0.004~nb for $\sigma^0_{\rm had}$.

\begin{table}[tb]
\renewcommand{\arraystretch}{1.2}
\begin{center}
\begin{tabular}{|l|r|r|}
\hline
 & $\Gamma_\PZ$ [GeV] & $\sigma^0_{\rm had}$ [nb] \\
\hline \hline
Born$\,+\,\OO(\alpha)$ & 2.49769 & 41.4687 \\
$+\OO(\alpha\as)$ & 2.49648 & 41.4758 \\
$+\OO(\alpha_\Pt\as^2,\,\alpha_\Pt\as^3,\,\alpha^2_\Pt\as,\,\alpha_\Pt^3)$ &
2.49559 & 41.4770 \\
$+\OO(N_f^2\alpha^2,N_f\alpha^2)$ & 2.49423 & 41.4884 \\
\hline
$-\OO(\alpha_\Pt\as^3)$ & 2.49430 & 41.4882 \\
\hline
Large-$\mt$ exp.\ for EW 2-loop & 2.49485 & 41.4840 \\
\hline
\end{tabular}
\end{center}
\vspace{-2ex}
\mycaption{Results for $\Gamma_\PZ$ and $\sigma^0_{\rm had}$, with $\MW$ calculated
from $G_\mu$ using the same order of perturbation theory as indicated in each
line. In all cases, the complete radiator functions ${\cal R}_{\rm V,A}$
are included. The last two lines show the comparison between the result based on
the full fermionic two-loop EW corrections and the large-$\mt$ approximation
\cite{ewmt4,ewmt2,xfit}. For consistency of the comparison, the small
$\OO(\alpha_\Pt\as^3)$ contribution has been removed in the next-to-last line,
since this part is also missing in the last line, but the three-loop corrections
or order $\OO(\alpha_\Pt\as^2,\,\alpha^2_\Pt\as,\,\alpha_\Pt^3)$
are included in both.
\label{tab:res2}}
\end{table}

The new results, including all corrections described above and the currently
most precise result for $\MW$ \cite{mwtot}, 
can be accurately described by the simple parametrization formula
\begin{align}
&X = X_0 + c_1 L_\PH + c_2 \Delta_\Pt + c_3 \Delta_{\as} + c_4 \Delta_{\as}^2
 + c_5 \Delta_{\as}\Delta_\Pt + c_6 \Delta_\alpha + c_7 \Delta_\PZ, \\
&L_\PH = \log\frac{\MH}{125.7\gev}, \quad
 \Delta_t = \Bigl (\frac{\mt}{173.2\gev}\Bigr )^2-1, \quad
 \Delta_{\as} = \frac{\as}{0.1184}-1, \nonumber \\
& \Delta_\alpha = \frac{\Delta\alpha}{0.059}-1, \quad
 \Delta_Z = \frac{\MZ}{91.1876\gev}-1. \nonumber
\end{align}
The coefficients are given by
\begin{align}
&X=\Gamma_\PZ\text{ [MeV]}: &X_0 &= 2494.24, &&c_1 = -2.0, &&c_2 = 19.7,
&&c_3 = 58.60, \\ &&c_4 &= -4.0, &&c_5=8.0, &&c_6 = -55.9, &&c_7 = 9267\,;
\nonumber \displaybreak[0] \\[1ex]
&X=\sigma^0_{\rm had}\text{ [pb]}: &X_0 &= 41488.4, &&c_1 = 3.0, &&c_2 = 60.9,
&&c_3 = -579.4, \\ &&c_4 &= 38.1, &&c_5=7.3, &&c_6 = 85.4, &&c_7 = -86027\,.
\nonumber
\end{align}
This formula approximates the full results to better than 0.01~MeV and 0.1~pb,
respectively, for the input parameters within the ranges $\MH = 125.7\pm 2.5\gev$, $\mt = 173.2\pm 2.0\gev$,
$\as=0.1184\pm 0.0050$, $\Delta\alpha = 0.0590 \pm 0.0005$ and $\MZ = 91.1876 \pm
0.0042 \gev$.


\vspace{\bigskipamount}
In summary, the complete fermionic two-loop electroweak corrections to the $Z$-boson decay
width and the cross-section for $e^+e^- \to \text{hadrons}$ have been computed
within the Standard Model. Compared to known previous calculations, the new
contributions lead to shifts of 0.6~MeV in $\Gamma_\PZ$ and 0.004~nb in $\sigma^0_{\rm
had}$, which are smaller then, but
of comparable order of magnitude as the current experimental uncertainties
(2.3~MeV and 0.037~nb \cite{pdg}). Therefore, the new results are important for robust
predictions of these quantities, while the remaining theory uncertainty is
estimated to be relatively small. It
mainly stems from the missing bosonic $\OO(\alpha^2)$ contributions and 
$\OO(\alpha^2\as)$, $\OO(\alpha^3)$ and $\OO(\alpha\as^2)$ corrections beyond the
large-$\mt$ approximation, leading
to the estimates $\delta \Gamma_\PZ \approx 0.5$~MeV
and $\delta\sigma^0_{\rm had} \approx 0.006$~nb \cite{long}.


\section*{Acknowledgments}

The author wishes to thank S.~Mishima for useful discussions and detailed
numerical comparisons.
This work has been supported in part by the National Science Foundation under
grant no.\ PHY-1212635.


\end{document}